# Learning the Stress-Strain Fields in Digital Composites using Fourier Neural Operator

Meer Mehran Rashid[1], Tanu Pittie[1], Souvik Chakraborty[2,3,*], N. M. Anoop Krishnan[1,3,*]

[1]Department of Civil Engineering, Indian Institute of Technology Delhi, Hauz Khas, New Delhi 110016, India
[2]Department of Applied Mechanics, Indian Institute of Technology Delhi, Hauz Khas, New Delhi 110016, India
[3]Yardi School of Artificial Intelligence, Indian Institute of Technology Delhi, Hauz Khas, New Delhi 110016, India.

*Corresponding authors: S. Chakraborty (souvik@iitd.ac.in), N. M. A. Krishnan (krishnan@iitd.ac.in)

**Abstract**

Increased demands for high-performance materials have led to advanced composite materials with complex hierarchical designs. However, designing a tailored material microstructure with targeted properties and performance is extremely challenging due to the innumerable design combinations and prohibitive computational costs for physics-based solvers. In this study, we employ a neural operator-based framework, namely Fourier neural operator (FNO) to learn the mechanical response of 2D composites. We show that the FNO exhibits high-fidelity predictions of the complete stress and strain tensor fields for geometrically complex composite microstructures with very few training data and purely based on the microstructure. The model also exhibits zero-shot generalization on unseen arbitrary geometries with high accuracy. Furthermore, the model exhibits zero-shot super-resolution capabilities by predicting high-resolution stress and strain fields directly from low-resolution input configurations. Finally, the model also provides high-accuracy predictions of equivalent measures for stress-strain fields allowing realistic upscaling of the results.

***Keywords*:** Digital Composites, Fourier Neural Operator, Stress-Strain prediction, Zero-Shot generalization, Super-resolution.

## 1. Introduction

The surging demands for high-performance materials with diverse functionalities necessitates accurate models for capturing the material response valid for a wide array of scenarios[1,2]. Driven by the objective to engineer materials with tailored properties, such as stronger, lighter and stiffer materials; researchers have resorted to combining multiple phases to arrive at a superior composite material[3] that outperforms its constituent phases. Thus, while designing composites, the phase composition and microstructure are tuned to produce a mechanically superior material with desired properties and behavior. Many such advanced materials (bio-inspired materials[4–8], meta-materials[9,10], architected materials[11–14]) have been introduced with enhanced properties and performance. However, traditional manufacturing methods are incapable of exploiting material microstructure for improving the design besides difficulties in combining base materials. To fully harness the potential of material response, manipulations at the microstructural level have shown promise. In this regard, additive manufacturing has emerged as a feasible solution leading us to complex microstructural composites with unprecedented mechanical performance[15].

In order to investigate the material behavior, various modeling methods at different length scales have been used such as finite element[15] (FE), molecular dynamics[16,17] (MD) or density functional theory[18] (DFT) simulations. Creating virtual models and subjecting them to different representative real-world settings is a prerequisite for understanding the behavior, design improvisation, and further development. However, the plethoric possibilities of material configurations make it almost impossible to navigate and arrive at the optimal design. Using the above-stated computational tools in conjunction with a brute force trial and error approach to analyse different geometries is not a feasible solution to optimize the design. Besides, these methods are exorbitantly expensive lacking the means to transfer knowledge of one simulation to another. To address these drawbacks, recent advances in machine learning (ML) offer new solutions which are cost effective, fast as well as offer transferability.

Recent breakthroughs in ML have led to versatile algorithms perfectly modeling the complex nature of different scientific problems. With the advent of GPU and TPU facilitated ML and the abundance of data available, the training time has drastically reduced paving way for highly advanced[19–21] models solving intricate problems. The availability of ever-growing datasets has accelerated the advancements in the ML resulting in bigger and more complex predictive models with limitless parameters[22] that exhibit immense expressive power. The promise shown by the ML methods has led researchers from different domains to embrace and employ ML in their respective fields. For material science, ML models have provided a cheap alternative to resolve difficult challenges and achieve high fidelity results ergo facilitating computationally sophisticated research work. The application of ML techniques, particularly deep learning (DL) models, has facilitated novel material designs, accelerated material discovery, material modeling and property predictions[23–27].

Many studies[28–31] have focused on predicting the mechanical properties of different materials. By using two convolution neural network (CNN) framework-based architectures-SCSNet (single-channel stress prediction neural network) and StressNet to encode structure, boundary condition and external forces; Ni et. al[32] predicted the von-mises stress fields for 2D elastic cantilever structures. Sun et al.[33] used StressNet to predict the stress field in 2D slices of segmented tomography images of a fiber-reinforced polymer specimen. Yang et al.[34] combined principal component analysis (PCA) with CNN to predict the stress-strain behavior of binary composite over the entire failure path. Liu et al.[35] predicted micro-scale elastic strains in 3D voxel-based microstructure volume element. Sepasdar et al.[36] formulated a CNN-based framework to estimate the post-failure full-field stress distribution and crack pattern for carbon fiber-reinforced polymer composite. Similarly, Bhaduri et al.[37] considered U-Net architecture to map fiber configurations to von-mises stress fields. CNN becomes a natural choice when the solution is the image representation of any quantity due to its inherent capacity to detect local and global patterns. However, other DL based networks such as recurrent neural networks (RNNs) and generative models have also been utilized to estimate the mechanical response of materials. Mozaffar et al.[38] used RNNs to predict the plastic behavior of composite representative volume element (RVE). Various studies suggest generative models[39–41] while addressing the inverse problem of finding the potential material based on target properties. Furthermore, Yang et al.[1] used condition generative adversarial network GAN (cGAN) to predict the stress-strain fields for random two-phase microstructures. Besides, the results are used to derive secondary material properties. In a different work, the authors[2] use cGAN to predict the multiple tensorial stress-strain components. However, most of the models suffer in generalization, thereby, failing to make predictions for the input settings unseen to the model. While predicting stresses or strains, the existing studies predict a single tensorial component. Even though Yang et al.[2] predicted multiple components but each tensor element is predicted by a different trained model thereby, making it computationally expensive to predict a full tensor. Additionally, such pixel-to-pixel learning-based methods are incapable of resolving higher resolution inputs unseen during model training.

To address these drawbacks, we use Fourier neural operator[42] (FNO) to predict component-wise stress and strain for two-phase composites. Using the microstructure of the material alone as an input, we predict the normal and shear components of stress and strain tensor field in an end-to-end fashion. The model learns the relation between the design geometry and material response with high accuracy. By predicting the stress and strain tensors, the model learns the constitutive relation purely from data, devoid of any knowledge of the underlying physics of the problem. We demonstrate the ability of the ML model to generalize to unseen geometries with arbitrary shapes. Also, the super-resolution feature of the FNO model allows high-resolution output for low-resolution inputs. Using the stress and strain predictions, we also show that equivalent stress and strain-based quantities, *viz*, von-mises stress and equivalent strains can be estimated with high-accuracy allowing upscaling of the results to higher length scales.

## 2. Methodology

### 2.1 Dataset Preparation: Geometry, material properties, FE Modelling

Mode-I tensile test FE simulations are run on an 8 mm × 8 mm 2-D plate in ABAQUS to generate the initial dataset for the FNO. We use an arbitrary composite material made up of two individual components, namely soft material and stiff material. The modulus ratio ($E_{stiff}/E_{soft}$), as well as the toughness ratio ($G_{stiff}/G_{soft}$) for the two materials, is 10. However, the failure strain for both materials is kept equal. The square plate was divided into equal cells and each cell was assigned a material property (soft or stiff) randomly using a python script. However, the fraction of soft and stiff units is equal for all the FE samples. Therefore, each 2D composite has a material resolution of 8 × 8 and the overall image resolution is 48 × 48. Each pixel corresponds to a finite element in the FE configuration. The loading is applied in the horizontal direction (global y-direction) and the pre-crack is along the x-direction. Around 1500 distinct configurations are generated and used for simulating mode-I failure for the composite plate using the above method. These configurations are randomly generated to explore a wide range of design arrangements. In addition to this, we generate multiple test sets with varied material and image resolution discussed in the results section. All the results are post-processed using ABAQUS's python interface which includes extracting nodal and elemental information (type of material, displacement, strain, stress etc.) from the FE simulations used as part of the training and testing of the FNO model.

### 2.2 ML model and its training

#### 2.2.1 Model Description

Neural Operators (NOs): Traditionally, neural networks have been used to learn the mappings between finite-dimensional Euclidean spaces. For such network constructs, we can only feed discrete inputs to learn the underlying relation under a typical supervised learning setting. Recently, a new paradigm has been established known as the neural operator[43–45] to learn the mappings between infinite dimension Euclidean spaces. This generalization of neural network helps in learning the operator that maps input function space to solution space. We use these neural operators to solve the PDEs by specifically learning the operator that maps the input parameters $a \in \mathcal{A}$ to the solution space $u \in \mathcal{U}$. Let $D \subset \mathbb{R}^d$ be bounded and open set and $\mathcal{A} = (D; \mathbb{R}^{d_a})$ is the input function space, $\mathcal{U} = (D; \mathbb{R}^{d_u})$ is the output function space. $\mathcal{A}$ and $\mathcal{U}$ are the two Banach spaces of functions defined on domain D taking values in $\mathbb{R}^{d_a}$ and $\mathbb{R}^{d_u}$ respectively. $\mathcal{G}: \mathcal{A} * \theta \mapsto \mathcal{U}$ is the mapping that satisfies the PDE. Considering samples $\{a_j, u_j\}$ where $a_j$ is an independent and identically distributed (i.i.d) sequence sampled from the probability measure $\mu$ in $\mathcal{A}$ and $u_j = \mathcal{G}(a_j)$, the neural operator approximates the mapping $\mathcal{G}_\theta$ by minimising the following stated problem using the cost function $C: \mathcal{U} \times \mathcal{U} \mapsto \mathbb{R}$

$$\min_\theta E_{a \sim \mu}[C(\mathcal{G}_\theta(a), \mathcal{G}(a))] \qquad (1)$$

For the problem framework, we assume point-wise evaluations of both the input function $a_j$ and solution function $u_j$. Let $D_j = \{x_1, x_2, \ldots, x_m\}$ be the $m$ point discretization and $a_j$ and $u_j$ be the finite samples of input-output pairs accessible. In this computational setup, we work with these finite $m$ pair data $\{a_j, u_j\}_{j=1}^m$ to learn the non-linear differential operator $\mathcal{G}_\theta$ which approximates the $\mathcal{G}: \mathcal{A} \mapsto \mathcal{U}$ satisfying the governing PDE.

Fourier Neural Operator: Using data-driven and physics-informed neural networks to satisfy the differential operator has significantly sped up the solution convergence in contrast to the classical PDE solvers. However, these approaches become computationally expensive as they can only be trained for a single instance of PDE parameters $a \in \mathcal{A}$ and we require a different model training if the parameter setting is altered. To overcome this, the recently introduced Fourier Neural Operator is able to learn the non-linear differential operator which in turn learns the family of PDEs corresponding to different parameter values. Fourier Neural Operator is a state-of-the-art neural operator which can model a wide array of problems pertinent to the fields of fluid mechanics[42] and climate modelling[46]. The architectural breakdown of FNO is shown in Fig. 1. The input $a(x, t)$ is lifted to a higher dimension by fully connected shallow neural network P as $v(x); v(x) = P(a(x,t))$. This higher dimensional output is fed concurrently to an iterative setup of Fourier layer and convolution layer denoted as $v_{j+1} = \mathcal{H}(v_j) \; \forall \; j = 1, \ldots, T$ steps on $v_o(x)$. This typical update step is defined as

$$v_{j+1}(x) \coloneqq \sigma\left(W v_j(x) + (\mathcal{K}(a; \phi)v_j)(x)\right) \qquad \forall x \in D \qquad (2)$$

where, $\sigma(\cdot): \mathbb{R} \mapsto \mathbb{R}$ is a non-linear activation function, $W: \mathbb{R}^{d_v} \mapsto \mathbb{R}^{d_v}$ is a linear transformation, $\mathcal{K}: \mathcal{A} \times \theta \mapsto \mathcal{L}(\mathcal{U}, \mathcal{U})$ is the non-local integral operator. FNO treats $\mathcal{K}(a; \phi)$ to be a kernel integral transformation parametrized by $\phi \in \Theta_k$ k. This kernel integral operator is defined as:

$$\left(\mathcal{K}(v_j)(x)\right) = \int_D \kappa(a(x, y), x, y; \phi) v_j(y) dy \qquad x \in D, j \in [1, T] \qquad (3)$$

where $\kappa_\phi: \mathbb{R}^{2d+d_a} \mapsto R^{d_v \times d_v}$ is a neural network parametrized by $\theta \in \Theta$. It can be considered as the kernel function that is learned from the input data. By letting $\kappa(x, y) = \kappa(x - y)$, FNO replaces this kernel integral operator with a convolution operator defined in Fourier space where it is reduced to a basic multiplication operation. Let $\mathcal{F}$ denote the Fourier transform and $\mathcal{F}^{-1}$ the inverse Fourier transform, therefore (3) changes to

$$\left(\mathcal{K}(v_j)(x)\right) = \mathcal{F}^{-1}(\mathcal{F}(\kappa_\phi) \cdot \mathcal{F}(v_j))(x) \quad x \in D \qquad (4)$$

On parameterizing the $\kappa$ directly by its Fourier coefficients, we get

$$\mathcal{K}(v)(x) = \mathcal{F}^{-1}\left(R_\phi \times \mathcal{F}(v_j)\right)(x) \quad x \in D \qquad (5)$$

where $R_\phi$ is the Fourier Transform of periodic function $\kappa$. On assuming $\kappa$ as periodic, FNO exploits this by working with discrete Fourier modes of the Fourier expansion and truncates the series expansion at the maximum number of modes $\kappa_{max}$. The higher modes which are usually responsible for finer features are dropped to improve upon the speed of convergence as well as regularization. It is followed by an inverse Fourier transform to transform back to the spatial domain. The output of these iterative layers is fed to another shallow fully connected neural network which projects the data back to the target dimension. FNO takes advantage of the Fast Fourier Transform FFT algorithm to calculate the $\mathcal{F}$ and $\mathcal{F}^{-1}$ thereby responsible for its tremendous speed.

### 2.2.2 Model hyperparameters

The FNO architecture used for this study as described in Fig.1 comprises 6 layers in total. This includes 2 linear layers; one at the start and the other at the end having 32 and 128 nodes respectively and 4 Fourier layers in between. We train FNO by retaining the different number of modes and 12 modes are found to give the best results considering model accuracy and time needed during training. The model is trained on a single NVIDIA V100 GPU with 16GB memory using the PyTorch[47] framework. We use a smoother version of ReLU namely GELU (Gaussian cumulative distribution function) activation function, ADAM optimizer which is a first order gradient-based method to train 300 epochs with a batch size of 20. We keep the weight decay as 0.005 and the initial learning rate is fixed at 0.001 and it halves after every 100 epochs. During the training, we use an L$_2$ based loss function which is defined as:

$$\mathcal{L}_2 = \frac{\sqrt{\sum_{i=1}^{m}(u(x_i)-\hat{u}(x_i))^2}}{\sqrt{\sum_{i=1}^{m}(u(x_i))^2}} \tag{6}$$

where $u(x_i)$ is the ground truth and $\hat{u}(x_i)$ are the pixel-wise model prediction for the *i-th* point. The dataset is divided into 1200 training samples and 200 test samples. The above-stated values of batch size, learning rate, number of epochs and training set size have been considered after performing hyperparameter optimization.

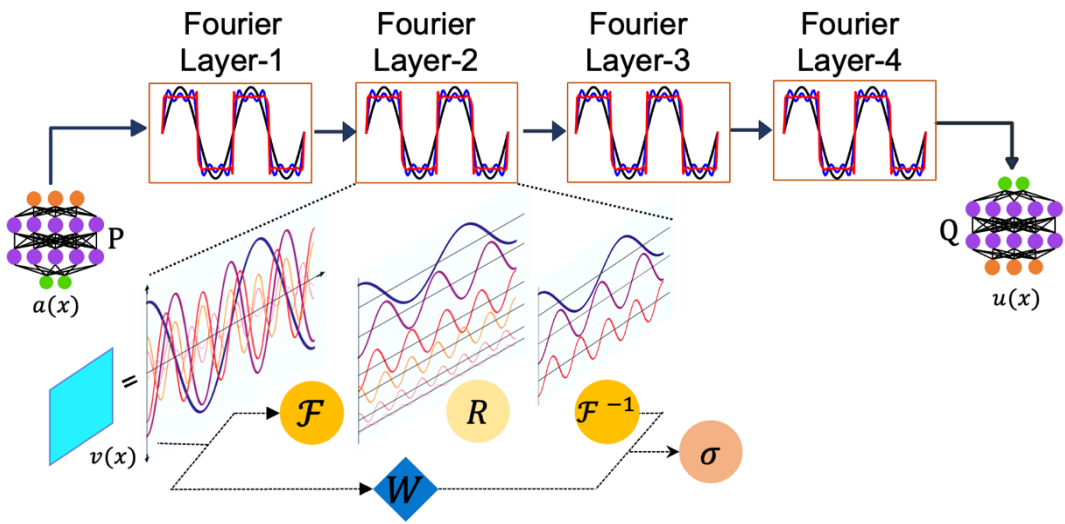

**Figure 1. Fourier Neural Operator-FNO network architecture**. The input is lifted to the higher dimensional channel space through a neural network P. The output of this linear layer is fed iteratively to the 4 Fourier Layers. Each Fourier Layer is an integral convolution in Fourier space. Taking the Fourier Transform $\mathcal{F}$ of the input $v(x)$ followed by a linear transformation $R$ on the lower modes and truncating higher modes; then applying inverse Fourier Transform $\mathcal{F}^{-1}$. Besides, the input is concurrently supplied to the local linear transformation $W$. The combined output of the spectral layers and convolution layer is acted upon by a non-linear activation function $\sigma$. Finally, neural network $Q$ projects the output back to the target dimension. $u(x)$ is the solution prediction of the FNO.

### 2.2.3 Evaluation Metrics

The output of the FE simulations is the element-wise data for each material geometry. The trained models are used to obtain the components of stress and strain tensors for material geometries having a unique configuration of soft and brittle units in their composition. Besides field variables, we evaluate global properties such as von-mises stresses and equivalent strain using the available field outputs. We present pixel to pixel comparison of the solutions obtained from the FNO model with those obtained from the numerical solver. To quantitatively assess the performance of the FNO model, we measure element-wise absolute error (AE) maps and relative error (RE) for the field variables defined as:

$$AE := \delta_{AE} = |\hat{u}(x_i) - u(x_i)| \quad (7)$$

$$RE := \delta_{RE} = \left|\frac{\hat{u}(x_i) - u(x_i)}{u(x_i)}\right| \quad (8)$$

where $\hat{u}(x_i)$ is the predicted value and $u(x_i)$ is the actual value for the *i-th* element. In addition to these, for every component, we plot the FEM vs prediction results for stress and strains along with specific cross-sectional directions. Among a wide range of available colour schemes, the colour spectrum used for plotting the results works best in terms of viewing the details at the soft and brittle interface as well as near the crack region.

### 3. Results

### 3.1 FNO framework

First, we briefly discuss the FNO-based framework used to predict the non-linear stress-strain response for the 2D hierarchical composite. Figure 2 presents the graphical workflow followed in this study. We consider a binary composite, that is, a composite consisting of two arbitrary materials of different stiffness. The initial geometry is generated in a chequered pattern with each square randomly assigned one of the two materials. Ground truth is generated using finite element (FE) simulations (see Methods for details). The randomly generated geometric configurations are given as input to the FNO model to predict the stress-strain response of the material by operator learning in a supervised fashion, where the ground truth is extracted from the FE simulations.

Specifically, we use the FNO to learn the constitutive relation for the digital composite by predicting the component-wise stress and strain fields. FNO belongs to the recently established neural operator class of deep learning frameworks that are used to model a wide range of complex problems (mainly governed by PDEs) e.g. turbulent flows, multiphase flow, weather predictions. The parameters are learned in the Fourier space where the output of each Fourier layer is truncated by dropping higher Fourier modes mainly responsible for details of the construction. Broadly, FNO comprises of a lifting layer, iterative kernel integration layers or the Fourier layers and the projection layer. The input is lifted to the higher dimension using a lifting layer P, essentially a linear layer with 32 nodes in our case. The higher dimensional output goes through an iterative setup of Fourier layers and within each Fourier layer the physical representation is convoluted with the kernel function which amounts to simple multiplication in Fourier space. FNO utilizes the FFT algorithm to transform both the entities followed by product operation. The output is filtered by removing the higher modes thereby neglecting the high frequency noise in the feature information. This filtration leads to the model speed up as well as model generalization. FNO uses the inverse FFT to transform back these filtered modes to the spatial domain. Finally, the output of these Fourier layers is projected to the target dimension using the projection layer Q which is a linear layer with 128 nodes. Further details of FNO used in the present work are provided in the Methods section.

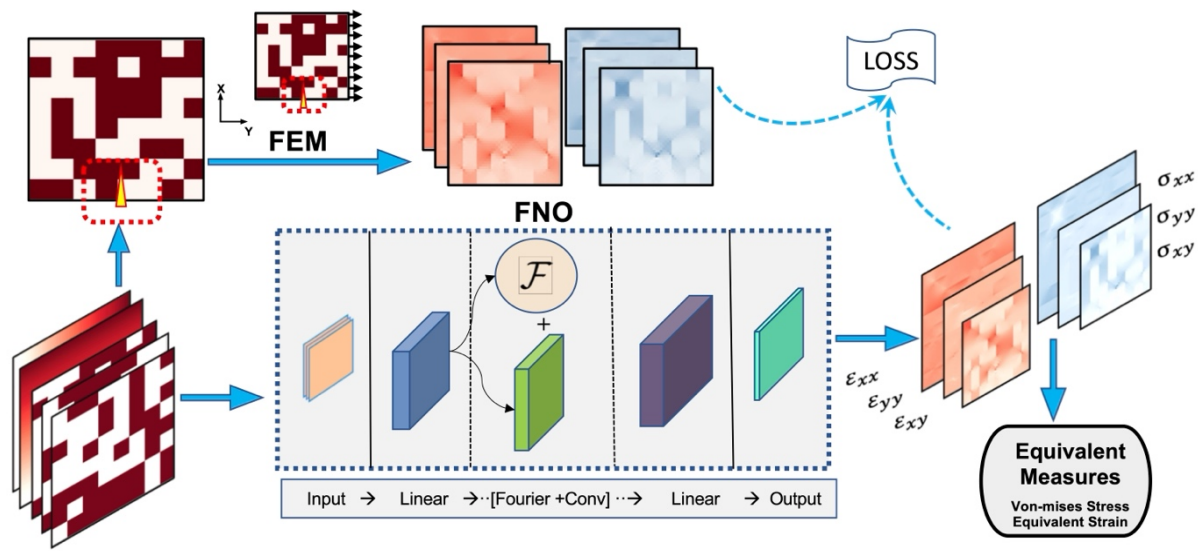

**Figure 2. Workflow.** The 2D digital composite geometry is analyzed for the mode-I tensile test using FEM. Pre-crack is along the x-direction and loading is applied in the y-direction. This simulation is done to establish the ground truth for model training. Material geometry image is the input to the FNO model. FNO framework used has 6 layers; 2 linear layers and 4 Fourier Layers. The model is trained separately to predict stresses and strains but for each of these field variables, all the components are predicted in a single pass. The trained model outputs are validated against the accurate FEM results besides testing it for unseen geometries. The tensor components are used to derive scalar-valued equivalent measures such as von-mises stress and equivalent strains.

### 3.2 Stress-strain prediction

To demonstrate the ability of FNO to predict complex stress-strain patterns, we train the model with the data on mode-I quasistatic fracture response of digital composite (see Fig. 2) having soft and stiff units. For each material geometry, the model predicts three stress components, viz., $\sigma_{xx}$, $\sigma_{yy}$ and $\sigma_{xy}$ as well as three strain components $\varepsilon_{xx}$, $\varepsilon_{yy}$ and $\varepsilon_{xy}$. The FNO based stress-strain predictions for a typical composite are shown in Fig. 3. Unlike previous studies[2], our model predicts full stress-strain tensors in a single pass. Instead of training different models for each component, we just train two models to predict all the components of the stress and strain tensor. The model input is the material geometry image having $8 \times 8$ material units and $48 \times 48$ image resolution. Figure 3(a) reveals the component-wise stresses compared to the

FEM-based ground truth. The output field maps for each component are also 48 × 48. The predicted strain fields qualitatively and quantitatively agree with the ground truth except in the regions of the crack tip and a rare occurrence at the soft-stiff unit interface. It is important to mention that calculating a single value-based error for the whole image doesn't provide insight into the accuracy of the results. In order to quantitatively evaluate the accuracy of the model predictions, we calculate the pixel-wise absolute error (AE) and absolute relative error (RE) fractions for each component. Only a few pixels show relatively high error; this is due to the development of localized stress concentrations at and around such regions. For the rest of the grid points, the results are consistent with the FEM output and precisely capture the stress patterns for each component. As expected, relatively higher stress values are generated in stiffer units. It can be visualized clearly in the stress distribution maps, especially for the $\sigma_{yy}$ component, as the loading is applied in the y-direction. Likewise, Fig. 3(b) shows the three strain component predictions that are obtained from a different trained model. The strain field predictions exactly resemble the ground truth (FEM results). The exactness of global strain patterns for composite geometry is remarkable, especially the ability to pick up the crack tip position besides the complex response at soft-stiff unit interfaces. The difficulties in model predictions at the crack tip are expected since this represents a discontinuity that is even challenging for conventional solvers. Creating a DL framework with the capacity to exactly capture the crack behavior is a potential area for future work.

Now, we plot the results along cross-sections in two specific directions to further illustrate the accuracy of the model predictions for stress and strain components. We choose two lines *XX* and *YY* along the horizontal and vertical directions respectively and plot the model predictions vs ground truth (FEM) for each of the tensor components. Figure 4 shows the results for one such example wherein for the same material geometry, we show stress components $\sigma_{xx}$, $\sigma_{yy}$, $\sigma_{xy}$ in Fig. 4(a) and strain components $\varepsilon_{xx}$, $\varepsilon_{yy}$ and $\varepsilon_{xy}$ in Fig. 4(b). The results almost precisely match the ground truth, thereby capturing the complex nature of these quantities. Except the boundary and regions around crack tip, FNO yields precise results with predictions overlapping the FE results. This is valid for any component of stress or strain tensor evaluated by the model. With such accurate predictions, the model can be used to achieve high fidelity results for field quantities thereby enabling us to explore the design landscape as well as comprehensively understand the material behavior.

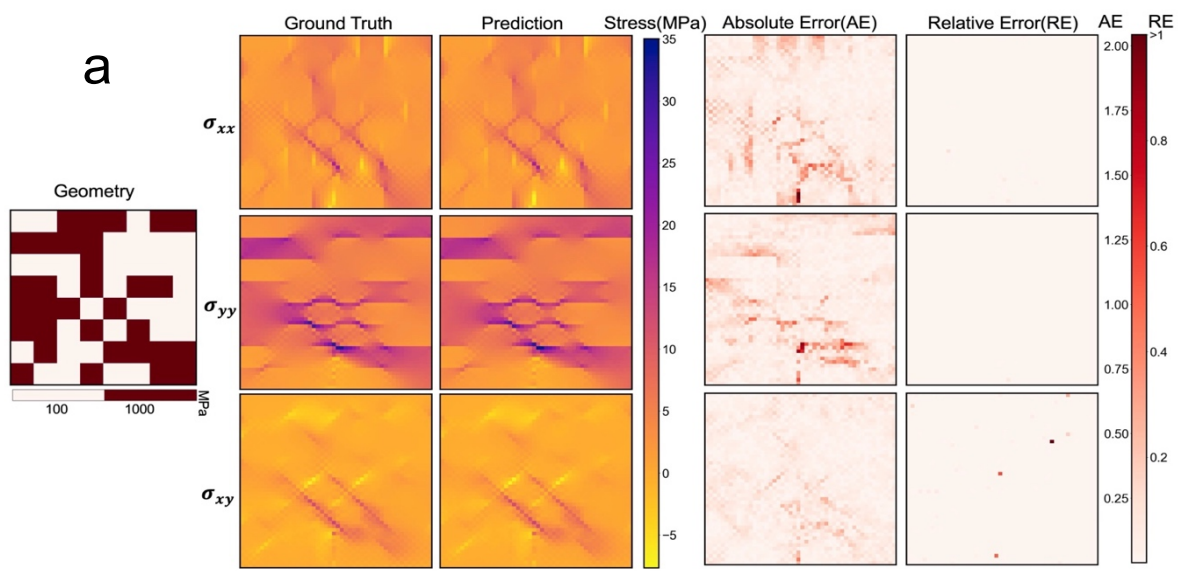

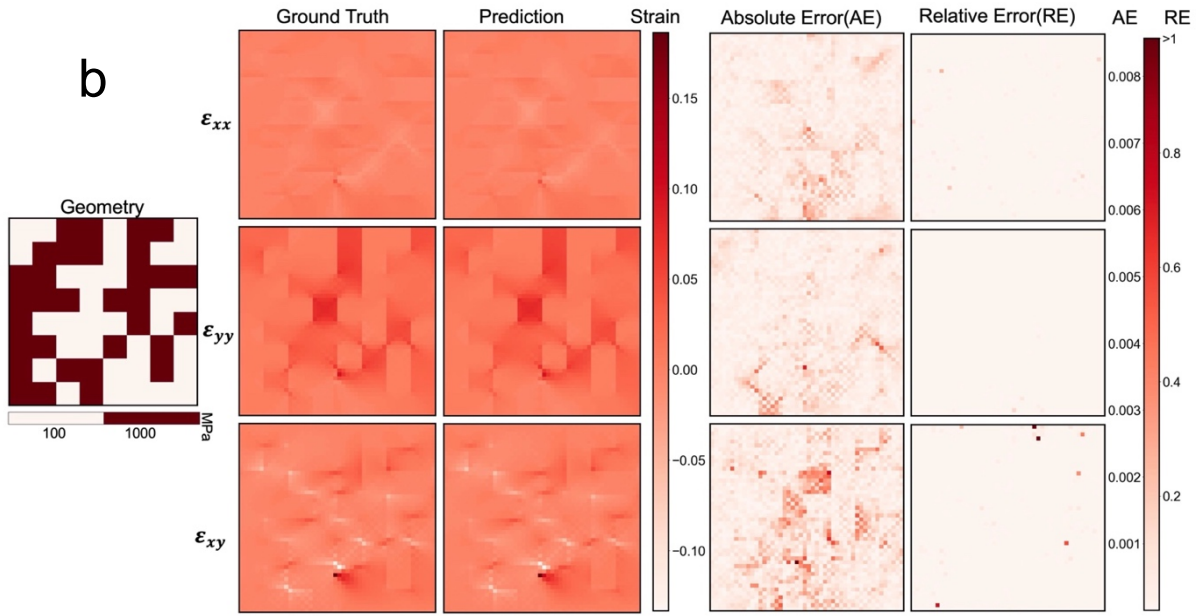

**Figure 3. Component-wise stress-strain prediction maps by the ML model for the 2D composite compared to high fidelity FEM solution.** (a) FNO predicting the three stress components $\sigma_{xx}$, $\sigma_{yy}$, $\sigma_{xy}$ for a typical composite material. The stress distribution is compared to the ground truth. Pixel-wise absolute relative error (AE) and absolute relative error (RE) maps are also shown corresponding to each stress component. (b) Similarly, strain components $\varepsilon_{xx}$, $\varepsilon_{yy}$ and $\varepsilon_{xy}$ predictions by the model compared with the FEM results. AE and RE maps for point-wise error quantification.

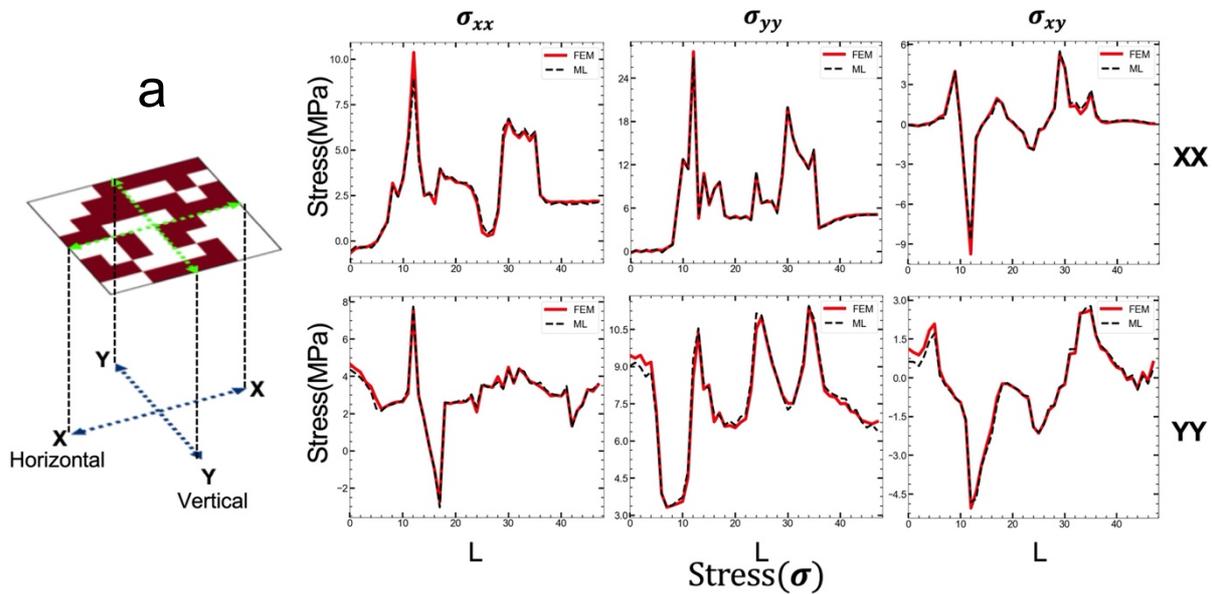

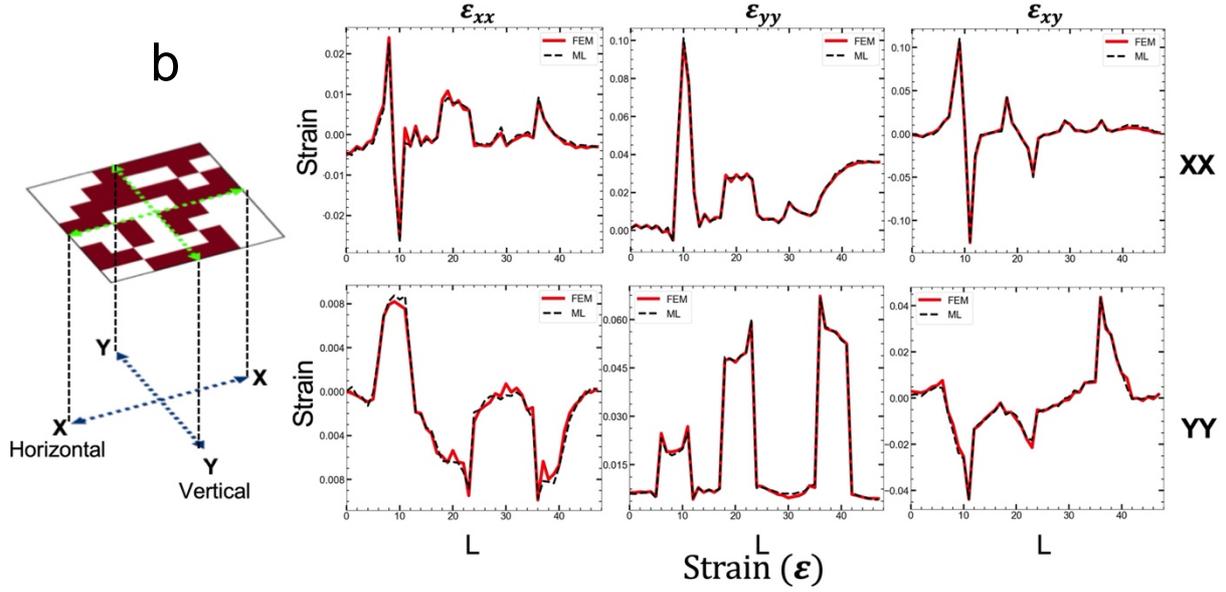

**Figure 4. Quantitative comparison of tensorial components along the specific cross-sectional directions.** (a) Comparison of stress values for each component along the *XX* and *YY* directions. (b) Comparison of strain values for each component along the *XX* and *YY* directions.

### 3.3 Material and Pixel-wise Super Resolution

Until above, the models have been trained on the input material geometry image of $8 \times 8$ grid of soft and stiff units and the overall image resolution for the input and output image maps was $48 \times 48$. The material resolution of $8 \times 8$ represents a simpler material configuration, but the geometries are more complex in real-world applications and a higher material grid resolution is observed. To address this challenge, we exploit the super-resolution capability of FNO, both in spatial as well as temporal domains. The model trained on lower resolution data can be evaluated for higher resolution making FNO discretization invariant. Unlike classical solvers whose results are significantly affected by the size of discretization, FNO is able to transfer the solution from lower resolution to higher resolution. This is possible because FNO by design learns the parameters, which are the Fourier modes in Fourier space. We exploit this feature and use the trained model to evaluate stress and strain fields on higher resolution images. For the case of material super-resolution, we test the model with input geometries having $16 \times 16$ material grid and $96 \times 96$ overall image resolution. The model predictions are shown in Fig. 5(a) depicting the $\varepsilon_{yy}$ component predictions *viz. a viz.* ground truth. Based on these results, we conclude that the model fairly captures the strain details for this high-resolution image as well as the crack tip position. The ability of the model to predict for higher material resolution shows its capability to predict at multiple length scales.

At times, we are interested in finer resolution details in the outputs, mainly around the stress concentrations or for the purpose of high-fidelity solutions. In classical solver setups such as FEM we are required to take small mesh sizes to capture small-scale details. However, this imposes severe computational costs and makes such analysis inefficient. FNO's ability to transfer solutions across different resolutions puts it in a unique list of frameworks available that feature super-resolution functionality with tremendous speed up. In this case, we test the model for inputs having fixed material grid size of $8 \times 8$ but varied overall image resolution. Figure 5(b) shows the vertical component of strain predicted for (i) $104 \times 104$ size and (ii)

200 × 200 size images along with absolute error plots. The model captures the strain patterns for both resolutions with decent accuracy. At points of stress concentration model suffers a bit because, in general, the deep learning models have the tendency to smoothen spikes (here stress concentrations) to lower the total loss. Overall, the ML model trained on lower resolution data can be used to fetch results for a finer domain discretization with acceptable accuracy. With such performance, the results of the ML based surrogate significantly reduce the costs of such analysis aimed at achieving high precision results. Since the FNO model utilizes larger Fourier modes for feature training, this leads to the loss of small details. Therefore, one can increase the number of Fourier layers without dropping any Fourier modes to capture more sharp details but at the cost of computational efficiency.

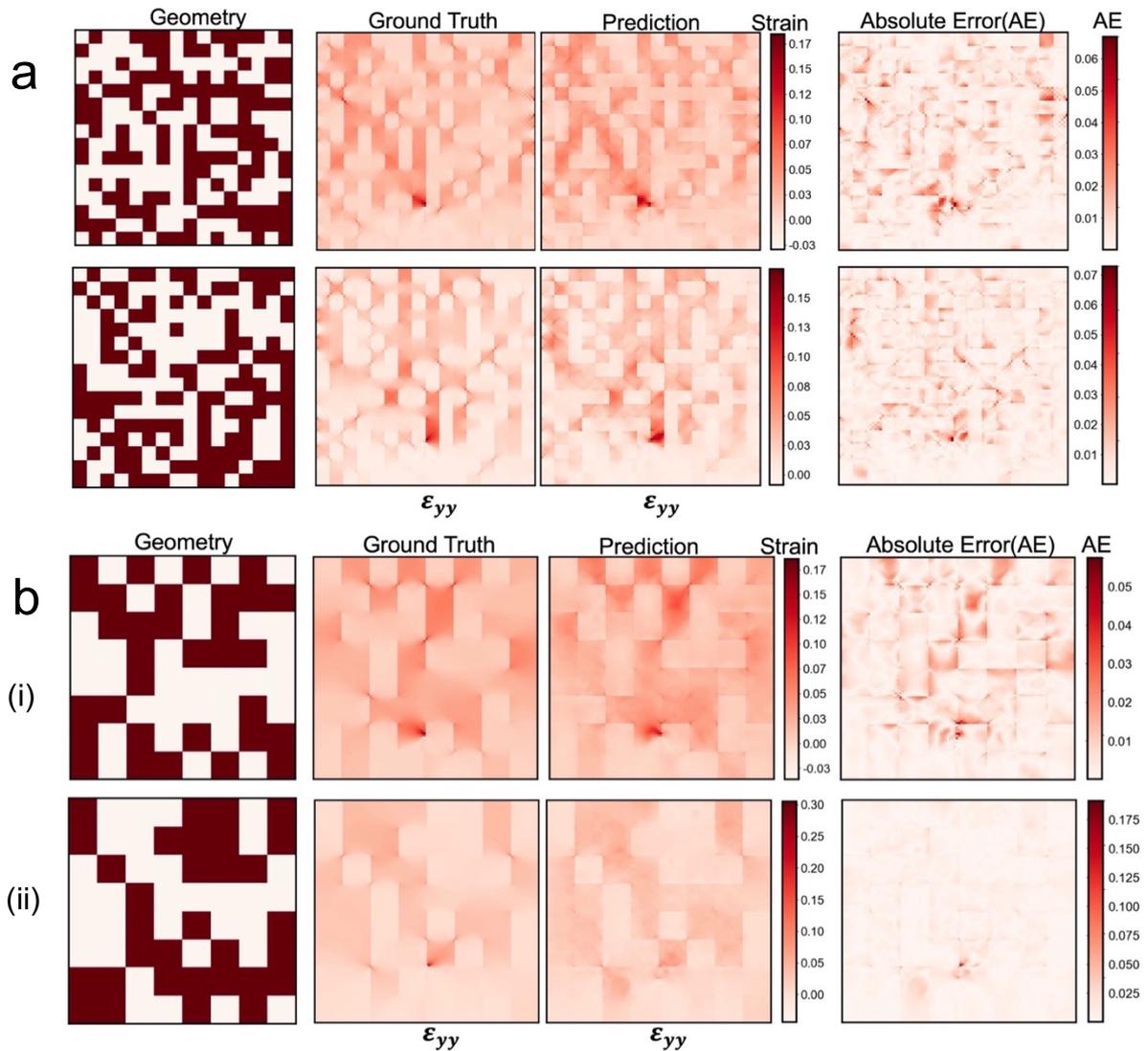

**Figure 5. Super-Resolution.** (a) Material-wise super-resolution: model trained on geometric configurations with 8 × 8 material grid tested for 16 × 16 material grid. Results are shown for two random geometric configurations along with AE distribution. Since loading is applied in y-direction, the model is trained to predict $\varepsilon_{yy}$ component (b) Pixel-wise super-resolution: The model trained on 48 × 48 image resolution is used to predict solution for higher resolution domain (i) 104 × 104 (ii) 200 × 200 image resolution. This establishes the robustness of the model with the ability to query solutions at new points in the domain.

## 3.4 Zero-shot Generalization to Unseen Geometries

The real-world geometries of the composites can get complex having any type of material distribution. Therefore, to extend the ambit of our model we test the model for arbitrary shapes. Earlier, the model has been trained on geometries with a chequerboard pattern of soft and stiff units; we now test it for geometries with arbitrary material distributions. These unseen geometries no longer have equal fractions of soft and stiff units which was the case during model training. To demonstrate such a possibility, we prepare a test set with random non-chequered material geometries with similar FE settings as mentioned in the Methodology section. The geometries are created to represent wide complexities possible in the design paradigm of such composites. Herein, we provide the results for three typical arbitrary geometries in Fig. 6. Since the composites are loaded in the horizontal direction (y-direction in FE setup), we evaluate the model for $\varepsilon_{yy}$ component. It is quite evident that the model generalizes well to composites having complex shapes exhibiting an extraordinary performance for zero-shot predictions. From the above results, it can be concluded, in principle, the model has been able to learn the complex mechanical behavior without being provided with any knowledge of the underlying physics/mechanics. We no longer need different models for component-wise field quantity evaluations or different conditions be it the changing geometry, changing soft-stiff unit fractions, or finer material resolution. The versatility of the ML model to predict for a wide range of scenarios can be used to optimize different mechanical properties previously computationally inaccessible.

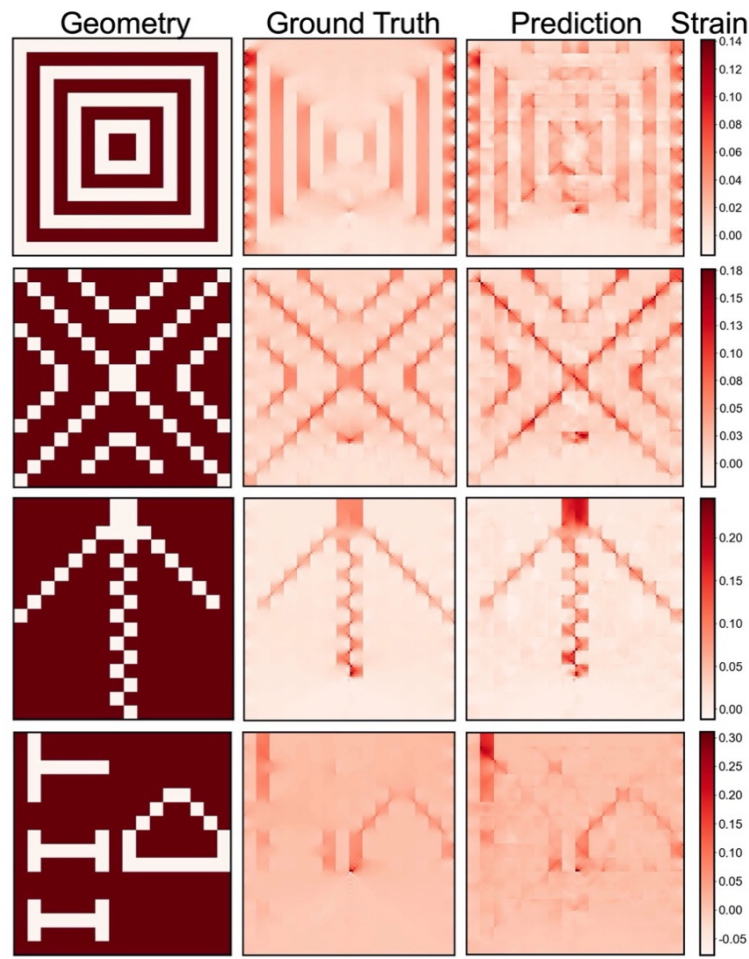

**Figure 6. Zero-shot prediction for non-chequered material geometries**. The model trained on chessboard geometry of soft and stiff units is tested against arbitrary non-chequered geometries with varying fractions of soft/stiff units. Direct comparison of $\varepsilon_{yy}$ values of ML model vs FEM shown for three typical examples.

### 3.5 Von-mises Stresses and Equivalent Strains

Until now, all the stress and strain components were directly outputted by the model. By using these results from the ML models, we aim to predict the equivalent quantities, specifically von-mises stress and equivalent strains. Von-mises stress is the non-tensorial measure of stresses calculated using normal and shear components. Likewise, equivalent strain is an effective single-valued measure of strain components. The available studies use end to end approach to calculate the equivalent quantities by training ML models to predict such quantities citing the complexities in predicting multiple tensorial components in a single pass. However, our model has the capacity to predict multiple components and hence, the equivalent measures of stresses and strains are computed by postprocessing the stress and strain tensor obtained using the trained FNO. Figure 7 shows the results for von-mises stress and equivalent strain distribution for a typical digital 2D composite compared with the results from FEM. On analyzing the results, we find that both the quantities closely match the ground truth (FEM results) suggesting the model's capacity to recognize complex material behavior. The robustness of the ML model to derive secondary mechanical quantities without the need to explicitly train for such can be used to evaluate the array of design settings leading to a composite with superior mechanical performance.

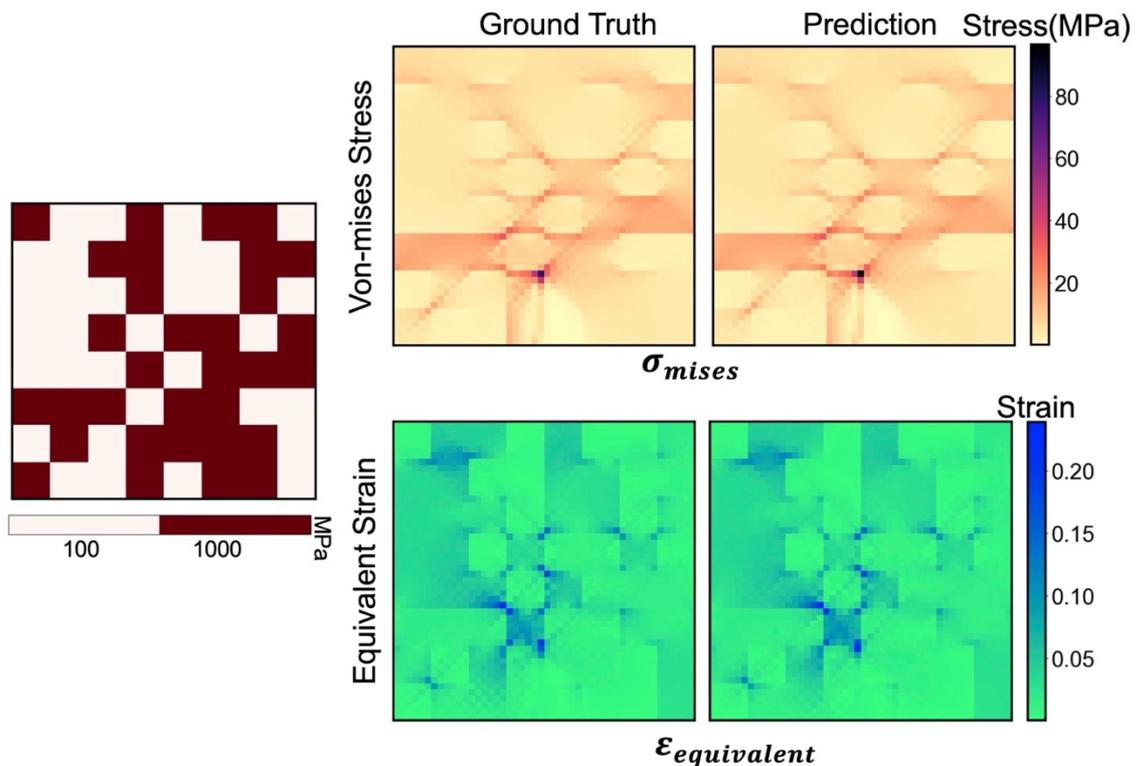

**Figure 7. Measurement of equivalent stress-strain quantities.** End to end approach is used for this study to predict stress-strain tensorial components. Using these results to derive the equivalent stress and strain measures viz, von-mises stress and equivalent strains. Distribution of these quantities is shown here for a typical geometry.

## 4. CONCLUSION

Altogether, in this study, we use a neural operator-based framework, namely, FNO to evaluate the mechanical response of digital composites subjected to tensile loading. To this extent, material geometries are randomly generated consisting of two distinct constituents of equal proportions. An end-to-end approach is used to predict the tensorial components of stress-strain fields. We show that the model trained on a fixed 48 × 48 resolution geometry images encoding material microstructure, exhibits excellent agreement with the ground truth obtained from the FE simulations. Further, we show that the FNO trained on a given resolution exhibits zero-shot generalisability to super-resolution both pixel-wise and material-wise. In addition, we show that the FNO exhibits zero-generalisability to complex geometries with varying percentage of the constituent materials. Finally, FNO also provides excellent predictions for the equivalent stresses and strains, allowing realistic upscaling of the results. These results substantiate the multifunctionality of the FNO model by generalizing over unknown microstructural shapes as well as outputting high-resolution predictions for low-resolution inputs.

At this juncture, it is worth mentioning some of the open challenges that remain to be addressed. Although the model provides excellent predictions in an overall fashion, the model exhibits inferior predictions for the stress concentrations at the crack tip, the improvement of which requires further work. Similarly, the crack propagation in the present work is modeled in a quasi-static fashion. Modeling dynamic fracture with varying time steps of integration remains an open challenge to be modeled in FNO. In addition, the ability of FNO to generalize to unseen boundary conditions remains to be explored. Finally, incorporating physics-based information to model the dynamics of crack propagation can be an interesting extension that can significantly enhance the performance of the model, while reducing the computational cost.

**Acknowledgments**

Authors thank the High-Performance Computing (HPC) facility at IIT Delhi for computational and storage resources.